\documentclass[%
 reprint,
superscriptaddress,
 amsmath,amssymb,
 aps,
prb,
]{revtex4-2}

\usepackage[english]{babel}
\usepackage{graphicx}
\usepackage{color}      
\usepackage{dcolumn}
\usepackage{bm}
\usepackage{amsmath,esint,mathtools}

\begin{document}

\preprint{APS/123-QED}
\title{Generalized electromagnetic theorems for non-local plasmonics}


\author{\'Emilie Sakat}
 \email{emilie.sakat@c2n.upsaclay.fr}
\affiliation{Universit\'e Paris Saclay, Center for Nanoscience and Nanotechnology, C2N UMR9001, CNRS, 91120 Palaiseau, France}
\author{Antoine Moreau}%
\affiliation{Universit\'e Clermont Auvergne, CNRS, Institut Pascal, F-63000 Clermont-Ferrand, France}%


\author{Jean-Paul Hugonin}
\affiliation{Universit\'e Paris-Saclay, Institut d'Optique Graduate School, CNRS, Laboratoire Charles Fabry, 91127 Palaiseau, France
}%



\begin{abstract}

The ultraconfined light of plasmonic modes put their effective wavelength close to the  mean free path of electrons inside the metal electron gas. The Drude model, which can not take the repulsive interactions of electrons into account, then clearly begins to show its limits. In an intermediate length scale where a full quantum treatment is computationally prohibitive, the semiclassical hydrodynamic model, instrinsically non-local, has proven successful. Here we generalize the expression for the absorption volume density and the reciprocity theorem in the framework of this hydrodynamic model. We validate numerically these generalized theorems and show that using classical expressions instead leads to large discrepancies.

\end{abstract}

\maketitle



Plasmonic mode volumes, which are several orders of magnitude smaller than the cubic wavelength, offer unique opportunities to produce local heating \cite{Ndukaife2016}, to enhance chemical reactions at a precise location \cite{Cui2015,Gelle2019} or to increase the spontaneous emission rate of a quantum emitter associated to a cavity \cite{Bozhevolnyi2016,Bozhevolnyi2017,Leng2018}. These deeply subwavelength volumes occur because light is noticeably slowed down when it propagates in the vicinity of metals, leading to slow guided modes \cite{ajib2019energy}. At these ultraconfined scales, the effective wavelength of the plasmonic guided modes gets close to the mean free path of free carriers inside the metal electron gas, so that the Drude model, which can not take such interactions into account, clearly begins to show its limits \cite{Ciraci2012,Moreau2013}.

A precise modeling of the light-matter interactions in these systems requires an accurate description of both quantum effects and far-field radiation. Time-dependent density functional theory (TD-DFT) \cite{Ciraci2016} or single-band theory in the Random-Phase Approximation (RPA) \cite{Delerue2017} can provide full first-principles quantum treatment. However, they become computationally
prohibitive for sizes that exceed a few nanometers and for high density of carriers. At the intermediate length scale, the semiclassical hydrodynamic Drude model (HDM) has proven successful to describe experimental results \cite{Ciraci2012,Raza2013,Christensen2014,Mortensen2014}. In the framework of this model, the optical response of metals cannot be described simply by a local permittivity any more so that they are called spatially dispersive or non-local.

Losses are both sought after for many applications of plasmonics like sensing, heating or even electro-optical modulation \cite{haffner2018low},
while being often responsible for the limitations of ultimately miniaturized plasmonic devices \cite{akselrod2014probing} -- in which case minimizing losses is crucial. It is thus of paramount importance to be able to compute accurately the local absorption inside nanostructures for which spatial dispersion (non-local effects) has to be taken into account. The standard expression for the local absorption in the harmonic regime

\begin{equation}\label{eq:alpha}
    \alpha_{\mathrm{abs}}(\mathbf{r},\omega) = \frac{\omega \mathrm{Im}(\varepsilon(\omega))|\textbf{E}(\mathbf{r},\omega)|^2}{2},
\end{equation}
where $\omega$ is the frequency, $\varepsilon$ the dielectric permittivity of the material and $\textbf{E}$ the total electric field, is obviously not valid anymore.

The lack of a correct formula has led many authors to look for a work-around or to rely only on far-field quantities, as it hinders from computing the actual absorption cross-section by integrating directly the absorption density \cite{David2011,Toscano2012,Moreau2013,Ciraci2013,Christensen2014,Mortensen2014,Toscano2015,Maack2017,Pitelet2018,Pitelet2019}. Furthermore, given the fact that the electric field in the framework of the hydrodynamic model contains a longitudinal component, the regular expression of the reciprocity theorem can not be relied on either. While, for other descriptions of non-local effects, the issue has been underlined and adressed \cite{Leung2008,Xie2009}, this has not been done for the widely used HDM model. This is a source of concern, as many other properties or laws are proven using this theorem (e.g. the Kirchhoff law), and it is routinely used to check the accuracy of numerical methods \cite{wakabayashi2018reciprocity}.

In this work, we first derive a formula for the absorption volume density that is valid even in a non-local medium. Then, we derive a generalized version of the reciprocity theorem valid also when one of the sources is located inside a non-local medium. The expression we obtain corresponds to the classical expression for the theorem except only the transverse component of the electric field should be considered. Finally, our theoretical results are validated numerically, using simulation methods which integrate the hydrodynamic model rigorously. 
These results show that different expressions (like Eq.\eqref{eq:alpha}) for the absorption volume density may lead to a large error when computing the absorption cross-section of a spherical nanoparticle.

\section{\label{sec:HDM} Hard-wall hydrodynamic model}

Drude's model is first based on the idea that the density of volume currents can be integrated as an effective polarization $\mathbf{P_f}$ of the medium, using the relation $\mathbf{j}=\frac{\partial \mathbf{P_f}}{\partial t}$. A metal can thus always be described with such a polarization. The following assumption is that the movement of electrons is dependent only on the local electric field and that the repulsion between electrons inside the metal can be neglected. Such an hypothesis holds only if the typical scale of the field variation is large in comparison to the mean free path of electrons. However, in plasmonics guided modes typically tend to present a high effective index and thus very short effective wavelength. Therefore, the local hypothesis does not hold any more.

The hydrodynamic model \cite{Ciraci2013} is the most simple possible way to take into account the repulsion between electrons, by integrating a pressure term in the description of the electron gas response, which leads to the following relation between the electric field and the effective polarization:

\begin{equation}\label{eq:euler}
   \frac{\partial^2 \mathbf{P_f}}{\partial t^2}+\gamma \frac{\partial \mathbf{P_f}}{\partial t} - \beta^2 \nabla(\nabla.\mathbf{P_f})=\varepsilon_0 \omega_p^2 \mathbf{E}.
\end{equation}
The electron pressure term  $\beta^2 \nabla(\nabla. \mathbf{P_f})$, coming from the
Thomas--Fermi theory of metals, obviously includes spatial derivatives of the polarization, making the description non-local. Let's underline here that still, the HDM model falls within the continuum theory, which considers that every mass point is composed by enough atoms and molecules to associate to this point macroscopic properties as a temperature T(r), a pression p(r), or in our case an absorption $\alpha_{abs}$(r). All these quantities are averages in the sense of statistical physics.
Let's note also that, even in the framework of the hydrodynamic model, the quantity $\chi_f=-\frac{\omega_p^2}{\omega^2+i\gamma\omega}$, which represents the local susceptibility of the Drude model plays an important role in the following.

The parameter $\beta$ quantifies the non-local effects and is called the hydrodynamic parameter. It can be defined as $\beta=\sqrt{3/5}\,v_F$ with $v_F=\frac{\hbar}{m^*}(3 \pi^2)^{1/3}n_0^{1/3}$, the Fermi velocity \cite{Ciraci2013}.

The bound electrons contribute to the response of the metal, however, their response can be considered as purely local and the corresponding polarization written
$\mathbf{P_b}=\varepsilon_0 \mathbf{\chi}_b \mathbf{E}(\mathbf{r})$. We can thus write the nullity of the divergence of the displacement field:
\begin{equation}\label{eq:D}
    \nabla.\mathbf{D}=\varepsilon_0 \nabla.\mathbf{E} + \nabla.\mathbf{P_f} + \nabla.\mathbf{P_b}=0
\end{equation}
in order to express and inject $\nabla.\mathbf{P_f}$ in Eq.\ref{eq:euler}. In the harmonic regime, the following non-local expression for $\mathbf{P_f}$ is thus obtained:

\begin{equation}\label{eq:Pf}
    \mathbf{P_f}=\frac{-\varepsilon_0 \omega_p^2}{\omega^2+i\gamma\omega} \left[\mathbf{E}-\frac{\beta^2(1+\chi_b)}{\omega_p^2}\nabla(\nabla.\mathbf{E})\right].
\end{equation}

Then, Ampère's circuital law $\nabla\times\mathbf{H}=-i\omega \mathbf{D}=-i\omega(\varepsilon_0\mathbf{E}+\mathbf{P_f}+\mathbf{P_b})$ yields
\begin{equation}\label{eq:rotH}
    \nabla\times\mathbf{H}=-i\omega\varepsilon_0\left[(1+\chi_b)\mathbf{E}+\frac{\mathbf{P_f}}{\varepsilon_0}\right]=-i\omega\varepsilon_0\varepsilon\left[\mathbf{E}-\alpha\nabla(\nabla.\mathbf{E})\right]
\end{equation}
with $\alpha=\frac{\chi_f(\varepsilon-\chi_f)\beta^2}{\varepsilon \omega_p^2}$ and $\varepsilon$=$1+\chi_f+\chi_b$.

Let's note here that the permittivity $\varepsilon$ is the local bulk permittivity of the medium (case where no repulsion between electrons is considered). This is what allows us to make a link with the local description: when the parameter beta tends towards zero, nonlocality is supposed to disappear gradually and we retrieve the purely local model.

According to Helmholtz's theorem, any sufficiently smooth, rapidly decaying vector field in three dimensions can be resolved into the sum of an irrotational vector field (the longitudinal component) and a divergence-free vector field (the transverse component). 

In a non-local medium, $\mathbf{E}$ can always be written as:
\begin{align*} 
\mathbf{E} &=\mathbf{E}_0+\mathbf{e} \\ 
\mathbf{H} &=\mathbf{H}_0
\end{align*}
($\mathbf{E}_0$,$\mathbf{H}_0$) corresponding to a transverse wave and $\mathbf{e}$ to a longitudinal wave. By definition $\nabla\times\mathbf{H}_0$=$-i\omega\varepsilon_0\varepsilon\mathbf{E}_0$ and $\nabla.\mathbf{E}_0$=0. Thus, by using the second term of Eq.\ref{eq:rotH}, we can write in the non-local media:
\begin{align*}
\nabla\times(\mathbf{H}-\mathbf{H}_0)=0&=-i\omega\varepsilon_0\varepsilon[(\mathbf{E}-\mathbf{E}_0)-\alpha\nabla(\nabla.\mathbf{E})]\\
0&=-i\omega\varepsilon_0\varepsilon[\mathbf{e}-\alpha\nabla(\nabla.\mathbf{e})]
\end{align*}
If we now define $\rho$ such as $\mathbf{\rho}$=$\alpha\nabla.\mathbf{E}$, the above equation gives
\begin{equation}\label{eq:e_rho}
\mathbf{e}=\nabla\rho
\end{equation}
and given that $\nabla.\mathbf{E}$=$\nabla.\mathbf{E}_0$+$\nabla.\mathbf{e}$, another HDM fundamental relation is obtained:
\begin{equation}\label{eq:disp_rho}
\Delta\mathbf{\rho}=\frac{\rho}{\alpha}
\end{equation}

\begin{figure}[h!]
    \centerline{\includegraphics[width=0.6\columnwidth]{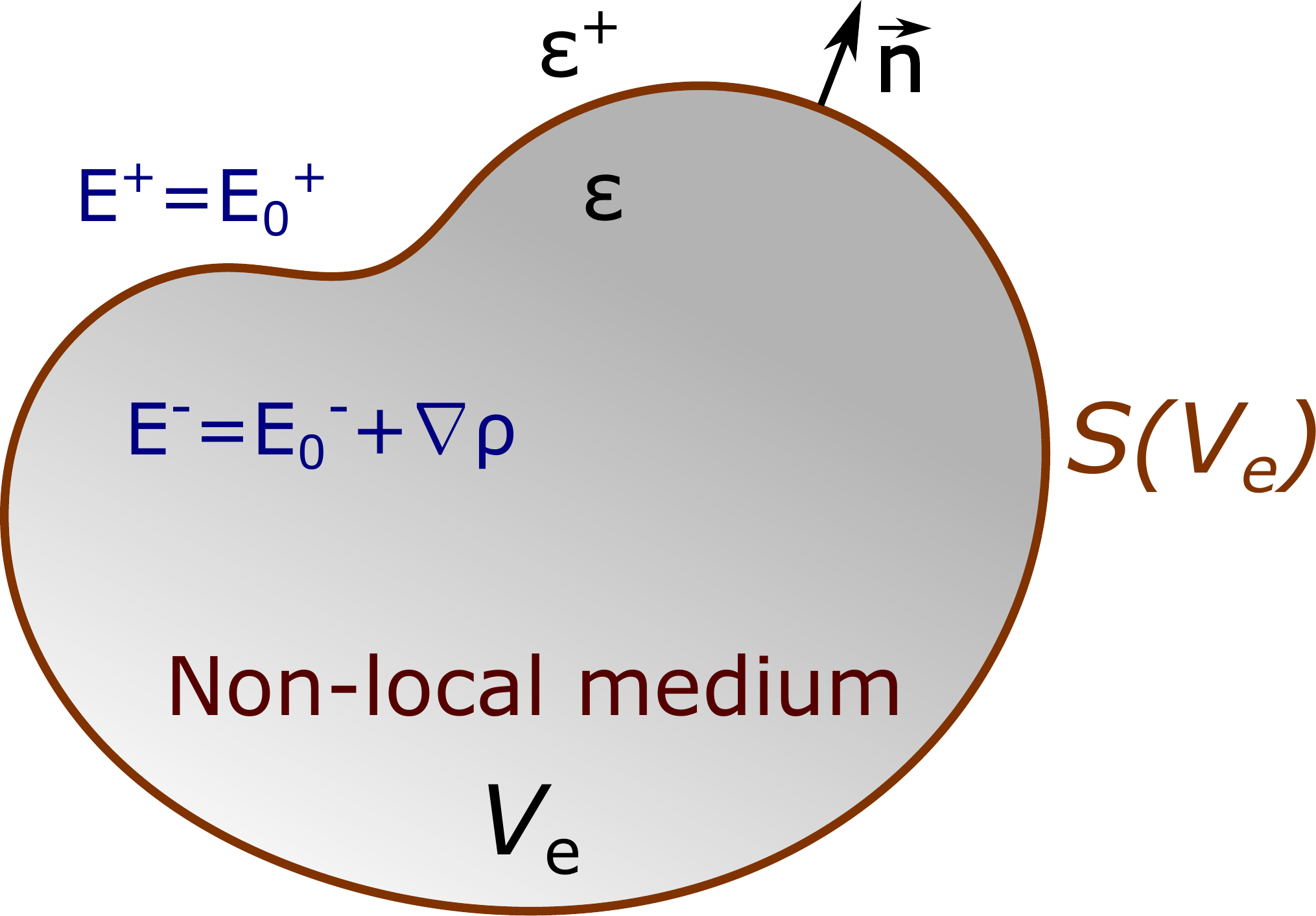}}
	\caption{Non-local medium of permittivity $\varepsilon$ inserted in a local background medium of permittivity $\varepsilon^+$ which is not necessarily spatially uniform.}
	\label{fig:principe}
\end{figure}

With these definitions, and in particular $\mathbf{e}=\alpha\nabla(\nabla.\mathbf{e})$, Eq. \ref{eq:rotH} leads to the following expression:
\begin{equation}\label{eq:new_eq5}
\frac{\mathbf{P}_f}{\varepsilon_0}=\chi_f\mathbf{E}-\varepsilon\mathbf{e}=\chi_f\mathbf{E}_0-(\varepsilon-\chi_f)\mathbf{e}
\end{equation}
In the usual implementation of the HDM, electron tunneling or electron density spill-out effects are neglected. This leads to the hard-wall boundary condition, stating that $\mathbf{P_f}.\mathbf{n}$=0 \cite{Moreau2013}. In this paper, we will limit ourselves to this hypothesis even if few works have proposed to go beyond \cite{Toscano2015,Ciraci2016}. Within this framework, the continuity conditions for the problem illustrated on Fig. \ref{fig:principe} are:\\
1. the tangential components of $\mathbf{E}$ and $\mathbf{H}$ are continuous,\\ 
2. the normal component of $\varepsilon\mathbf{E}_0$ is continuous,\\ 
3. the normal component of $(\varepsilon-\chi_f)\mathbf{E}$ is continuous.\\

These hard-wall boundary conditions along with Eq.\ref{eq:new_eq5} and Eq.\ref{eq:e_rho} imply one last fundamental relation of the HDM model at the non-local/local interface:
\begin{equation}\label{eq:grad_rho}
\nabla\mathbf{\rho}.\mathbf{n}=\frac{\chi_f}{(\varepsilon-\chi_f)}\mathbf{E}_{0}^-.\mathbf{n}
\end{equation}

This HDM model has been implemented in a home-made code based on a T-matrix formalism and treating multiple scattering in stratified media problems \cite{Bertrand2020}. In such code the incident and
scattered fields for an individual particle are decomposed on a complete and orthonormal basis of vector spherical harmonics (VSHs). The VSHs scattered by each particle are decomposed as incident fields on the other particles. The T-matrix connect the scattered field to the total incident field (incident light and scattered light of all other particles) of each particle. Let's note that with such code, the surface integral of the entering total Poynting Flux of each particle is automatically calculated. The implementation of the HDM model in this code is similar to what is done in Ref. \cite{Christensen2014,Maack2018}.

\section{\label{sec:PertesFormule}Non-local absorption volume density}

In order to describe the absorption power in the non-local medium with the hard-wall HDM model described above, we can write the surface integral of the Poynting vector flux around the non-local medium:

\begin{equation}\label{eq:PA}
    P_{\mathrm{abs}}(\omega) = -\frac{1}{2} \oiint_{S(V_e)} \mathrm{Re}(\mathbf{E^*}\times\mathbf{H}).\mathbf{n}.
\end{equation}

Given the continuity of the tangential components of the electromagnetic fields ($\mathbf{E},\mathbf{H}$), this integral can be calculated indifferently on the inner or on the outer contour of the non-local media. On the inner contour of the integral, Eq. \ref{eq:PA} gives
\begin{multline}\label{eq:PA_d}
    P_{\mathrm{abs}}(\omega) = -\frac{1}{2} \oiint_{S(V_e)} \mathrm{Re}(\mathbf{E_0^*}\times\mathbf{H}).\mathbf{n} \\
    -\frac{1}{2} \oiint_{S(V_e)} \mathrm{Re}(\mathbf{\nabla\mathbf{\rho^*}}\times\mathbf{H}).\mathbf{n}
\end{multline}

The first term of this equation is straightforward to express since $\mathbf{E_0}$ and $\mathbf{H}$ satisfy the classical Maxwell equation:
\begin{equation}\label{eq:PA_E0}
    -\frac{1}{2} \oiint_{S(V_e)} \mathrm{Re}(\mathbf{E_0^*}\times\mathbf{H}).\mathbf{n} = \frac{\omega\varepsilon_0}{2} \int_{V_e} \mathrm{Im}(\varepsilon(\omega))|\textbf{E}_0(\mathbf{r},\omega)|^2 d^3 \textbf{r} 
\end{equation}

Now let us consider the second term in Eq. \ref{eq:PA_d}. 
\begin{equation}\label{eq:PA_rho}
    \nabla.\left(\nabla\mathbf{\rho^*}\times\mathbf{H}\right)=-\nabla\mathbf{\rho^*}.(\nabla\times\mathbf{H})+(\nabla\times\nabla\mathbf{\rho^*}).\mathbf{H}
\end{equation}

By definition $\nabla\times\nabla\mathbf{\rho^*}$=0. Thus Eq. \ref{eq:PA_rho} reduces to:
\begin{equation}\label{eq:PA_rho2}
\nabla.\left(\mathbf{\nabla\mathbf{\rho^*}}\times\mathbf{H}\right)=\mathbf{\nabla\mathbf{\rho^*}.i\omega\varepsilon_0\varepsilon\mathbf{E}_0}
\end{equation}

If we now apply the Ostrogradski theorem (also called divergence theorem or Gauss's theorem) to express the second term of Eq. \ref{eq:PA_d}, we obtain:

\begin{equation}\label{eq:PA_rho3}
-\frac{1}{2} \oiint_{S(V_e)} \mathrm{Re}(\nabla\mathbf{\rho^*}\times\mathbf{H}).\mathbf{n}=-\frac{1}{2}\int_{V_e}\mathrm{Re}\left(\nabla\mathbf{\rho^*}.i\omega\varepsilon_0\varepsilon\mathbf{E}_0\right)
\end{equation}

Eq. \ref{eq:PA_d} can thus be rewritten as:
\begin{align}\label{eq:Formule}
P_{\mathrm{abs}}(\omega) & =\frac{\omega\varepsilon_0}{2}\left[ \int_{V_e} \mathrm{Im}(\varepsilon)|\textbf{E}_0|^2 -\int_{V_e}\mathrm{Re}\left[i\varepsilon\mathbf\mathbf{\nabla\mathbf{\rho^*}.{E}_0}\right]\right] \notag
\\& =\frac{\omega\varepsilon_0}{2}~\mathrm{Im}\left[\int_{V_e} \varepsilon\textbf{E}^*_0.\textbf{E}_0+\int_{V_e}\varepsilon\mathbf{\nabla\mathbf{\rho^*}.\mathbf{E}_0}\right] \notag
\\& =\frac{\omega\varepsilon_0}{2}~\mathrm{Im}\left[\int_{V_e} \varepsilon\textbf{E}^*.\textbf{E}_0\right]
\end{align}

giving a non-local expression for the absorption volume density:
\begin{equation}\label{eq:alpha_NL}
    \alpha_{\mathrm{abs}}(\mathbf{r},\omega) = \frac{\omega\varepsilon_0}{2}~\mathrm{Im}\left( \varepsilon\textbf{E}^*.\textbf{E}_0\right)
\end{equation}

This generalized expression is different from the local one given Eq. \ref{eq:alpha}. But if $V_e$ contains only local media, no longitudinal component exists, $\textbf{E}$=$\textbf{E}_0$ and the two expressions are equivalent. One important point to notice with this generalized expression of the absorption power is that unlike Eq. \ref{eq:alpha}, it can be negative. To understand the necessary conditions on the material properties to maintain the integral positive, this generalized expression can be rewritten differently (see demonstration in Appendix A). 
\begin{equation}\label{eq:Formule2}
P_{\mathrm{abs}}(\omega)=\frac{\omega\varepsilon_0}{2}~\int_{V_e}\left(\mathrm{Im}(\varepsilon-\chi_f)|\textbf{E}|^2+\mathrm{Im}(\chi_f)\left|\frac{\mathbf{P}_f}{\varepsilon_0\chi_f}\right|^2\right)
\end{equation}
This second expression highlights clearly that if $\mathrm{Im}(\varepsilon-\chi_f)$ and $\mathrm{Im}(\chi_f)$ are positive, which is in the overwhelming majority of cases true for metals, this generalized expression for the absorption power remains positive. We underline also that under this form, the \textit{total} absorption can be clearly attributed to the dielectric background and to currents inside the electron gas. One has to be careful with this second formula though. Integrands of Eq. \ref{eq:Formule} and Eq. \ref{eq:Formule2} are indeed not equivalents. The demonstration of the second one includes the condition $\textbf{P}_f$.$\textbf{n}$=0, which is true only on the non-local/local interface, while the demonstration of Eq. \ref{eq:Formule} can be done on any closed volume inside the non-local media. Thus, the two integrals are equals only for the closed volume surrounded by the contour where $\textbf{P}_f$.$\textbf{n}$=0. For other closed volumes inside the non-local media, this is not the case. In fact, only the integrand of Eq. \ref{eq:Formule} is an absorption volume density: it is the unique expression giving an integral equal to the Poynting flux surface integral on any closed volume inside the non-local media.\\\\
\begin{figure}[h!]
    \centerline{\includegraphics[width=\columnwidth]{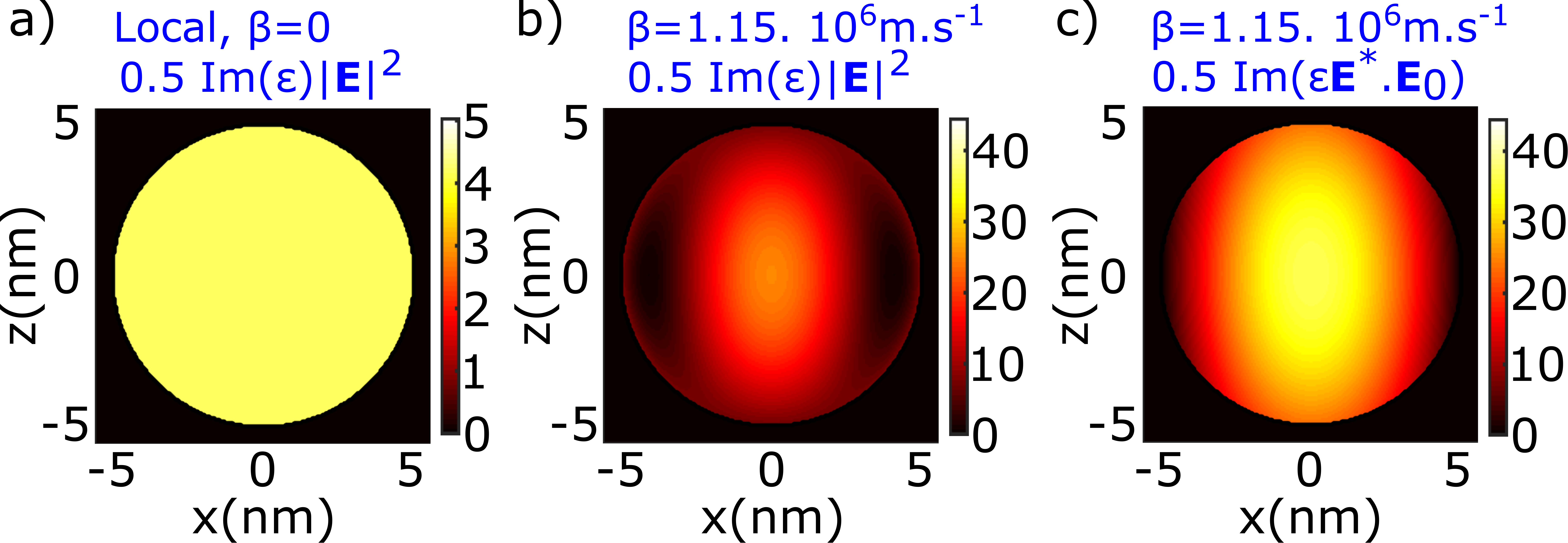}}
	\caption{Absorption volume density $\alpha_{\mathrm{abs}}(\mathbf{r})$ maps for a 5-nm radius nanosphere embedded in a background of refractive index n=1.5 at $\lambda$=2.4$\mathrm{\mu}$m. The HDM parameters for the nanosphere's permittivity are: $\varepsilon_\infty=1+\chi_b=4$, $\omega_p$=$1.7.10^{15}$rad.s$^{-1}$, $\gamma$= $1.52.10^{14}$rad.s$^{-1}$. a) classical Maxwell formula in the local case, i.e. $\beta=0$; b) classical Maxwell formula but the field \textbf{E}=$\textbf{E}_0$+\textbf{e} is derived within the HDM framework, $\beta$=$1.15.10^{6}$m.s$^{-1}$; c) formula \eqref{eq:alpha_NL}, $\beta$=$1.15.10^{6}$m.s$^{-1}$.  The absorption volume density is normalized by the norm of the Poynting vector of the incident plane wave. The sphere is illuminated by a plane wave polarized along the $x$-axis and propagating at normal incidence along the $z$ axis.}
	\label{fig:champ}
\end{figure}

Now, we illustrate the difference between the classical formula \eqref{eq:alpha} and the non-local expression \eqref{eq:alpha_NL} on an example. Figure \ref{fig:champ} compares absorption volume density maps calculated with these two formula for a 5-nm radius nanosphere embedded in a background medium of refractive index $n=1.5$. The permittivity chosen for the nanosphere is realistic and corresponds to a highly doped n-ITO nanocrystal. Figure \ref{fig:champ}a is calculated using Eq. \ref{eq:alpha} with the local Drude model. Figure \ref{fig:champ}b is calculated using Eq. \ref{eq:alpha}, but within the HDM framework, i.e. $|\mathbf{E}|=|\mathbf{E}_0+\nabla \rho|$. Figure \ref{fig:champ}c is calculated with Eq. \ref{eq:alpha_NL}. Figure \ref{fig:champ} clearly shows that while, in the local framework, electromagnetic field would be evaluated as quasi constant everywhere in the sphere (Fig. \ref{fig:champ}a), in the HDM framework on the contrary it is not the case, and if we compare the correct expression with the classical one (for both the field \textbf{E}=$\textbf{E}_0$+\textbf{e} is derived within the HDM framework), we see that absorption occurs closer to the surface of the metal with the correct formula (Fig. \ref{fig:champ}c) than with the classical one (Fig. \ref{fig:champ}b). In order to help understand theses maps, we put in Appendix B the intensity enhancement maps of the longitudinal and transverse fields, $|\mathbf{e}|^2$ and $|\mathbf{E}_0|^2$.

\begin{figure}[h!]
    \centerline{\includegraphics[width=\columnwidth]{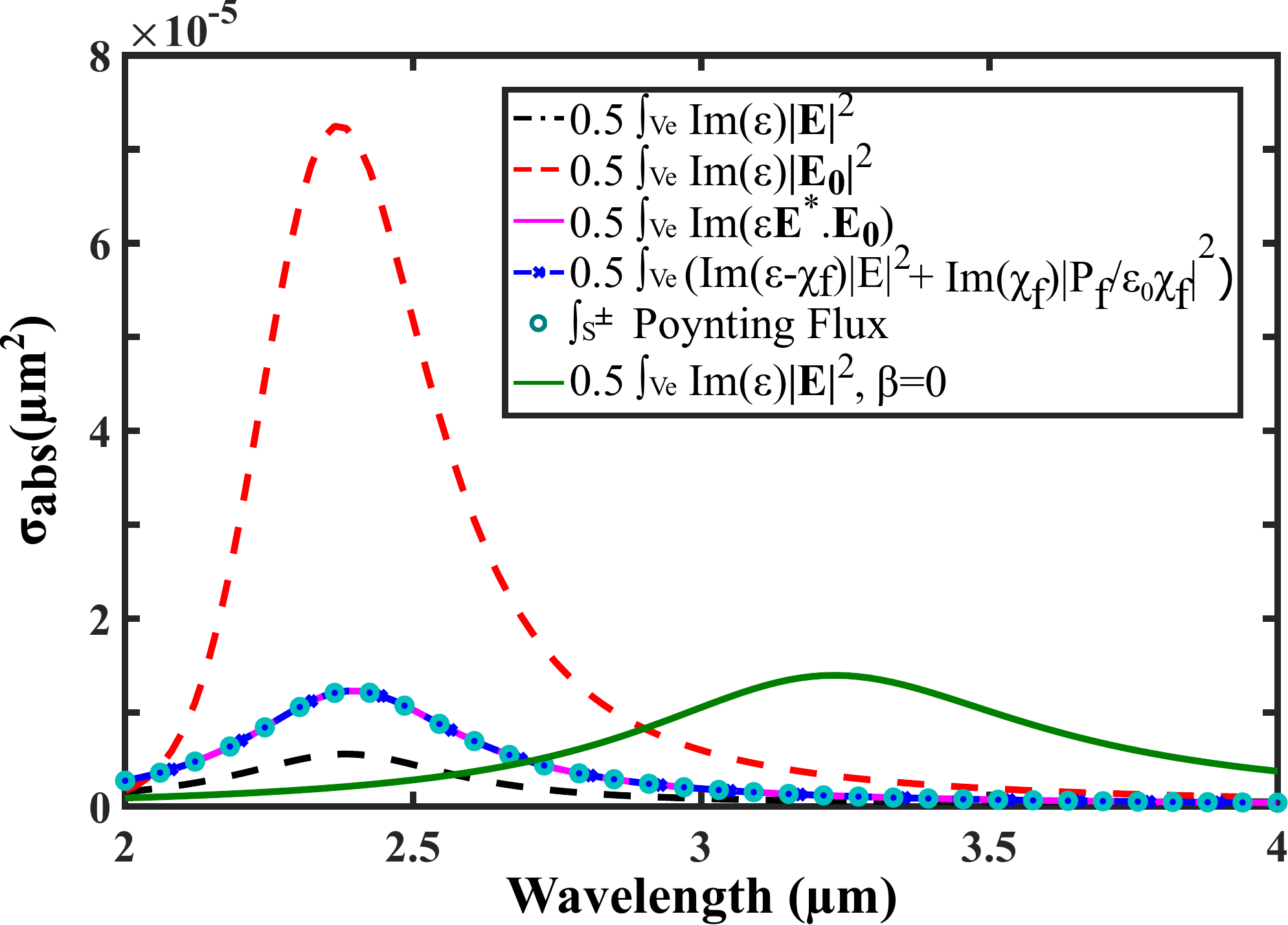}}
	\caption{Absorption cross-section as a function of the wavelength for the same non-local nanosphere as Fig.\ref{fig:champ}, assuming different expressions for the absorption volume density. Only the value obtained using Eq. \ref{eq:Formule} and Eq. \ref{eq:Formule2} matches the surface integral of the Poynting vector flux on the inner or outer contours surrounding the non-local medium. For comparaison, the classical local absorption cross-section spectrum is plotted in green.}
	\label{fig:spectre}
\end{figure}

Figure \ref{fig:spectre} represents the losses computed for the same 5-nm radius nanosphere by integrating different expressions for the absorption volume density $\alpha_{abs}$, as a function of the wavelength. Only Eq. \ref{eq:Formule} and Eq. \ref{eq:Formule2} allows to retrieve the correct value for the absorption cross-section computed using the flux of the Poynting vector. Assuming $\alpha_{abs}=\frac{1}{2}\,\mathrm{Im}(\varepsilon)|\mathbf{E}|^2$ yields a $\sigma_{abs}$ value more than twice as low as the actual value; choosing $\alpha_{abs}=\frac{1}{2}\,\mathrm{Im}(\varepsilon)|\mathbf{E}_0|^2$ gives a value almost six times too large. For comparison, we have also plotted the absorption cross-section calculated in the classical local case (where no repulsion between electrons is considered and so where $\beta$=0). We see that we retrieve the typical signature of the non-local effects, i.e. a blue-shift in the absorption cross-section spectrum. The effect here is very high because we have chosen a nanosphere with a permittivity close to the one allowing ideal absorption at $\lambda$=2.4$\mathrm{\mu}$m \cite{Grigoriev2015}. Moreover, the permittivity chosen is similar to the one of an highy-doped semicondutor (n-ITO), and it is now well-established that the non-local effects are expected to be more significant in highly-doped semiconductors than in noble metals and that they begin to occur for larger structures \cite{Maack2017,golestanizadeh2019hydrodynamic}. Indeed, the skin depth $\delta$ where the longitudinal component $\mathbf{e}$ produces a spatial variation in the induced charge density is proportional to $\frac{\beta}{\omega_p}$, itself proportional to $n_0^{-1/6}m^{*-1/2}$. In highly-doped semiconductors the free carrier density $n_0$ is orders of magnitude lower than in noble metals. The effective mass $m^{*}$ is also slightly lower. Thus the skin depth $\delta$ can be 5 times larger or more in highly-doped semiconductors. This optimized permittivity choice allows us to underline the difference between the local and non-local models, but this should not lead us to expect such a large impact of nonlocality is the norm.

We want to insist here that, while the quantity $\alpha_{\mathrm{abs}}(\mathbf{r},\omega)$ is strictly positive in a local medium, whenever bulk plasmons (the longitudinal wave), linked to nonlocality, are taken into account, losses may locally be negative. Below the plasma frequency, this occurs close to the surface, as the skin depth of bulk plasmons is much shorter than the skin depth of transverse waves. Above the plasma frequency, as bulk plasmons can propagate inside the medium, such a phenomenon can occur essentially anywhere. This could be particularly important for semiconductors since, unlike noble metals, no interband transition exists near their plasma frequency, and so the Drude model remains valid at the plasma frequency and above. We insist that, given our results and the link we have established with the flux of the Poynting vector, the global absorption would always be positive. This underlines how important this local absorption density may be to fully understand how light is absorbed by metallic or metallic-like media. 

\begin{figure}[h!]
    \centerline{\includegraphics[width=\columnwidth]{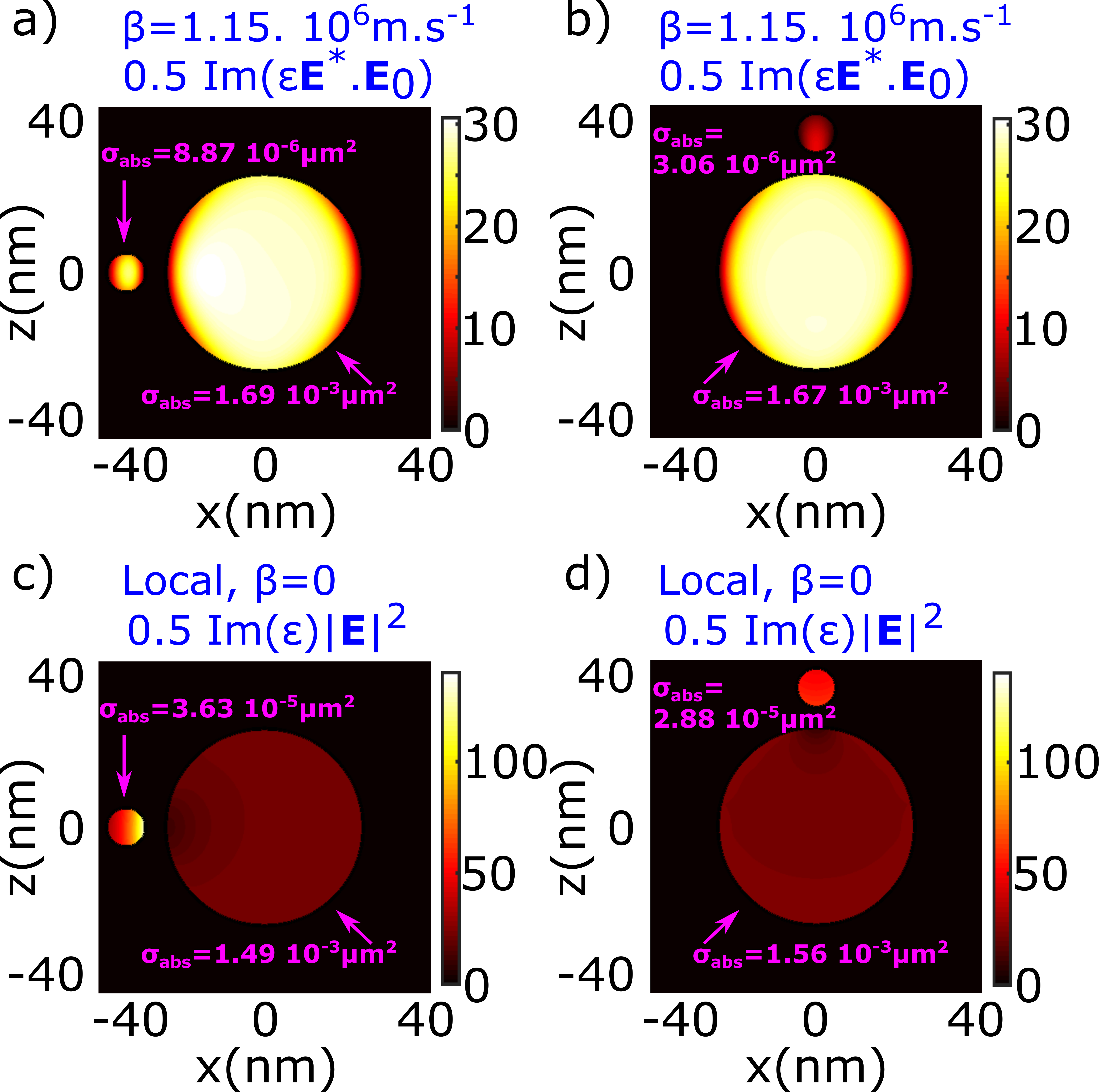}}
	\caption{Absorption volume density $\alpha_{\mathrm{abs}}(\mathbf{r})$ maps for a 5-nm radius nanosphere located in the vicinity of a 25-nm radius nanosphere. The gap between the two spheres is of 5 nm. The HDM parameters for the nanospheres permittivity are the same as Fig.\ref{fig:champ} and they are also embedded in a background of refractive index n=1.5. The wavelength is of $\lambda$=3.1$\mathrm{\mu}$m (resonance wavelength of the largest sphere). a) and b) $\alpha_{\mathrm{abs}}(\mathbf{r})$ follows formula \eqref{eq:alpha_NL} with $\beta$=$1.15.10^{6}$m.s$^{-1}$; c) and d) classical Maxwell formula in the local case, i.e. $\beta=0$. The absorption volume density is normalized by the norm of the Poynting vector of the incident plane wave. The total heating efficiency $\sigma_{abs}$ of each particles is indicated in magenta. The sphere is illuminated by a plane wave polarized along the $x$-axis and propagating at normal incidence along the $z$ axis.}
	\label{fig:champ_bis}
\end{figure}

This new expression is in particular the only way to estimate locally and accurately the losses in a non-local media. Such information can be very useful in practice to investigate and predict physical effects resulting from light absorption. In Ref. \cite{Baffou2009, Baffou2013} for example, the local absorption is used (in the classical local Maxwell case) to predict and compare the heating efficiency of different plasmonic nanoparticles (small, flat, elongated, or sharp nanoparticles). The efficiency differences between these particles are related to the capacity of the incoming electric field to penetrate inside the thin nanostructures and where it is occuring locally. Fig. \ref{fig:champ} highlights that the estimated power density of heat generation can be completely wrong if one uses a local version of Maxwell equations for a nanoparticle presenting non-local effects. Similarly, Fig. \ref{fig:champ_bis} illustrates a case where one searches to use a larger nanoparticle to heat or excite a nanoabsorber placed in its vicinity. Such methods are used for several applications in literature \cite{Baffou2013a, Sakat2018}. Here we give an example where a 25-nm-radius nanosphere heats a 5-nm-radius nanosphere. The obtained power density of heat generation is different with the present non-local formula than in the local case. Moreover, Fig. \ref{fig:champ_bis} shows that according to the nanoabsorber position with respect to the nanoantenna (the largest nanoparticle), the power density of heat generation changes rather heavily. The total heating efficiency of each particle obviously follows, and we see that while it is almost constant for the nanoantenna regardless of the position or the non-local character, for the nanoabsorber it can vary of a factor 3 according to the considered configuration (see Fig. \ref{fig:champ_bis}). The absorption volume density maps have been plotted here at $\lambda$=3.1$\mathrm{\mu}$m, which is the resonance wavelength of the largest sphere in the non-local case, but complete spectra of each particle absorption cross-section can be found in Appendix C. 

Given the importance of understanding where losses occur in any deeply subwavelength plasmonic structure, we think that expression (\ref{eq:Formule}) could prove very useful in the future by providing a more accurate physical picture of phenomena like quenching, thermoplasmics, or localized thermal emission \cite{anger2006enhancement,Baffou2013a,Sakat2018}.

\section{Reciprocity theorem with non-local media}

The reciprocity theorem allows to connect the electromagnetic fields generated by two different and arbitrary point sources.
By assuming time-harmonic fields in linear and local media in which the tensors $\varepsilon$ and $\mu$ are symmetric (reciprocal materials), the reciprocity theorem between two punctual time-harmonic source currents ($\mathbf{j}_1,\mathbf{j}_{m1}$) and ($\mathbf{j}_2,\mathbf{j}_{m2}$) located at positions $\mathbf{r}_1$ and $\mathbf{r}_2$ and which emit respectively the fields ($\mathbf{E}_1,\mathbf{H}_1$) and ($\mathbf{E}_2,\mathbf{H}_2$) can be written \cite{Carminati1998}: 
\begin{equation}\label{eq:rec}
\mathbf{j}_1\mathbf{E}_2(\mathbf{r}_1)+\mathbf{j}_{m1}\mathbf{H}_2(\mathbf{r}_1)=\mathbf{j}_2\mathbf{E}_1(\mathbf{r}_2)+\mathbf{j}_{m2}\mathbf{H}_1(\mathbf{r}_2).
\end{equation}

\begin{figure}[h!]
    \centerline{\includegraphics[width=\columnwidth]{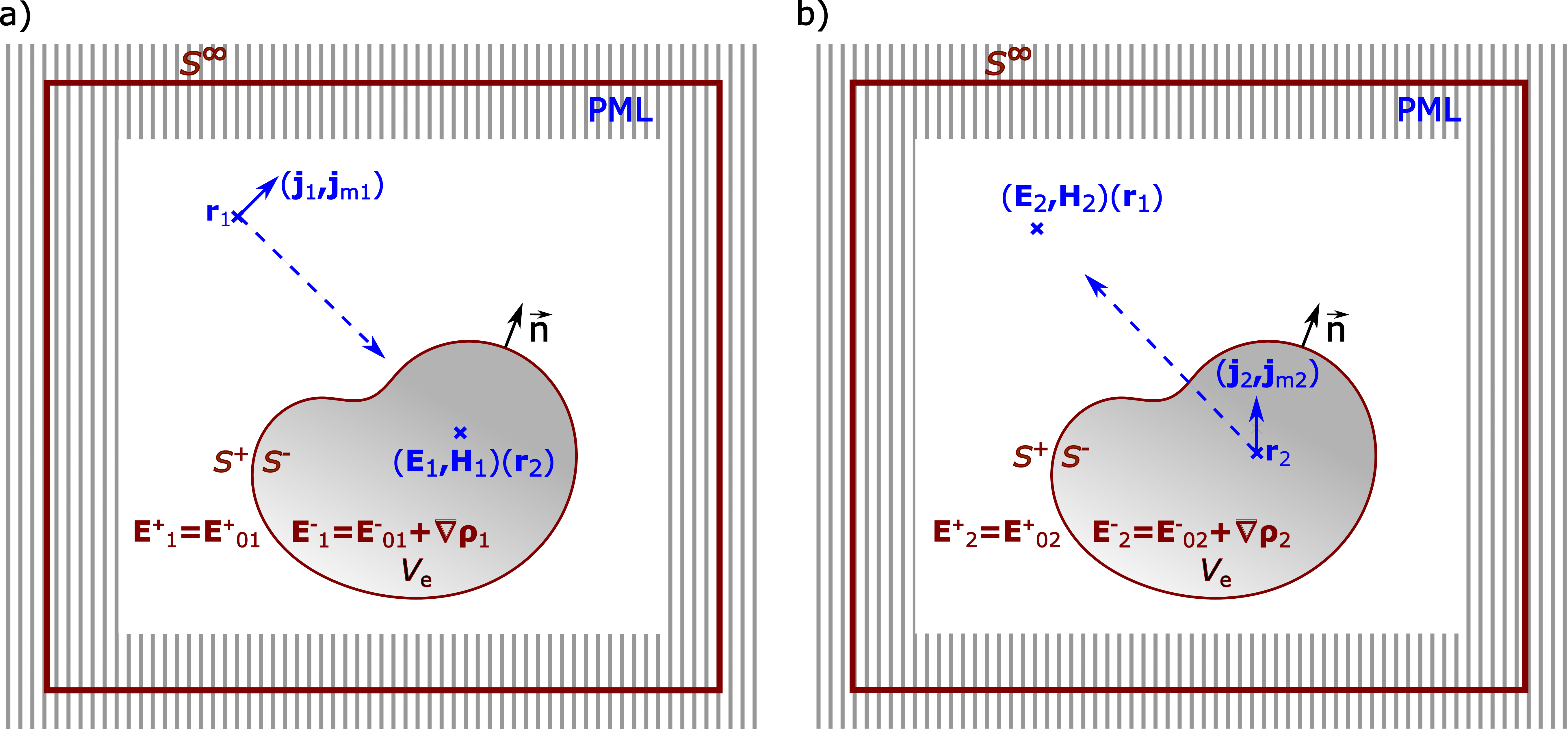}}
	\caption{A point source current ($\mathbf{j}_1$,$\mathbf{j}_{m1}$) located at a position $\mathbf{r}_1$ radiates, illuminating the non-local medium. It creates an electromagnetic field ($\mathbf{E}_1$,$\mathbf{H}_1$). (b) A point source current ($\mathbf{j}_2$,$\mathbf{j}_{m2}$) located at a position $\mathbf{r}_2$ inside the non-local medium radiates an electromagnetic field ($\mathbf{E}_2$,$\mathbf{H}_2$).}
	\label{fig:principe_rec}
\end{figure}

In the case where a non-local media is introduced, the reciprocity theorem can still be written but not under the same form. 
To find the right expression for the non-local reciprocity theorem, we consider the two problems described in Fig. \ref{fig:principe_rec}, where a source is located inside the non-local medium while the other is placed in the surrounding (local) medium. The Perfect Matched Layers (PML) around the object shown Fig. \ref{fig:principe_rec} are here only for numerical purposes,
in order to take into account the radiation boundary conditions at infinity.
Two solutions of Maxwell's equations in the HDM framework can thus be considered. In both case inside $V_e$:

\begin{equation}
\begin{cases}
 \nabla \times \mathbf{E}_k &=i\omega \mu \mu_0 \mathbf{H}_k + \mathbf{j}_{mk}\delta(\mathbf{r}-\mathbf{r}_k) \\
\nabla \times \mathbf{H}_k  &=-i\omega\varepsilon\varepsilon_0 \mathbf{E}_{0k} + \mathbf{j}_{k}\delta(\mathbf{r}-\mathbf{r}_k) \\
~~~~~~\mathbf{E}_k &=\mathbf{E}_{0k}+\nabla\mathbf{\rho}_k
\end{cases}
\end{equation}

with k=\{1,2\} and \{$\mathbf{j}_k,\mathbf{j}_{mk}$\} the electric and magnetic point sources respectively, simulated by dirac functions. 
 
The Lorentz reciprocity formula, which relates two time-harmonic solutions of Maxwell's equations can still be written on the transverse component $\mathbf{E}_{0}$ (a general form of Lorentz reciprocity formula and its derivation can be found for instance in Annex 3 of Ref. \cite{Lalanne2018}). Here we use this Lorentz reciprocity formula for the two transverse components $\mathbf{E}_{0}$ at the same frequency $\omega$, (i) on the closed surface surrounding the volume outside the non-local medium and located in between the contour $S^{\infty}$ located at infinity and the contour $S^+$ corresponding to the outer contour of the non-local medium, and (ii) on the closed surface $S^-$ surrounding the non-local volume $V_e$ (see Fig. \ref{fig:principe_rec}). The surface integral on $S^{\infty}$ being null by construction, this gives:
\begin{align*} 
-\oiint_{S^+}(\mathbf{E}_1\times\mathbf{H}_2-\mathbf{E}_2\times\mathbf{H}_1).dS\mathbf{n} &=\mathbf{j}_1\mathbf{E}_2(\mathbf{r}_1)+\mathbf{j}_{m1}\mathbf{H}_2(\mathbf{r}_1)\\
\oiint_{S^-}(\mathbf{E}_{01}\times\mathbf{H}_2-\mathbf{E}_{02}\times\mathbf{H}_1).dS\mathbf{n} &=
\begin{multlined}[t]
\shoveleft{-\left[\mathbf{j}_2\mathbf{E}_{01}(\mathbf{r}_2)\right.}\\
\shoveright{\left.+\mathbf{j}_{m2}\mathbf{H}_1(\mathbf{r}_2)\right]}
\end{multlined}
\end{align*}
We can now use the field continuity conditions to relate these two surface equations. It is in particular possible to show that (see Appendix D):
\begin{multline}\label{eq:int_annex}
\oiint_{S^+}(\mathbf{E}_1\times\mathbf{H}_2-\mathbf{E}_2\times\mathbf{H}_1).dS\mathbf{n} =\\
\oiint_{S^-}(\mathbf{E}_{01}\times\mathbf{H}_2-\mathbf{E}_{02}\times\mathbf{H}_1).dS\mathbf{n} 
\end{multline}
which immediately leads to the non-local reciprocity theorem:
\begin{equation}\label{eq:rec_NL}
\mathbf{j}_1\mathbf{E}_2(\mathbf{r}_1)+\mathbf{j}_{m1}\mathbf{H}_2(\mathbf{r}_1)=\mathbf{j}_2\mathbf{E}_{01}(\mathbf{r}_2)+\mathbf{j}_{m2}\mathbf{H}_1(\mathbf{r}_2) 
\end{equation}

Note that in the non-local medium, unlike Eq. \ref{eq:rec}, the reciprocity theorem is valid only for the transverse component $\mathbf{E}_0$ of the field and not for the total field $\mathbf{E}$. As a consequence, the Green tensor is symmetrical for $\mathbf{E}_0$ but not for the total field $\mathbf{E}$.

\begin{figure}
    \centerline{\includegraphics[width=\columnwidth]{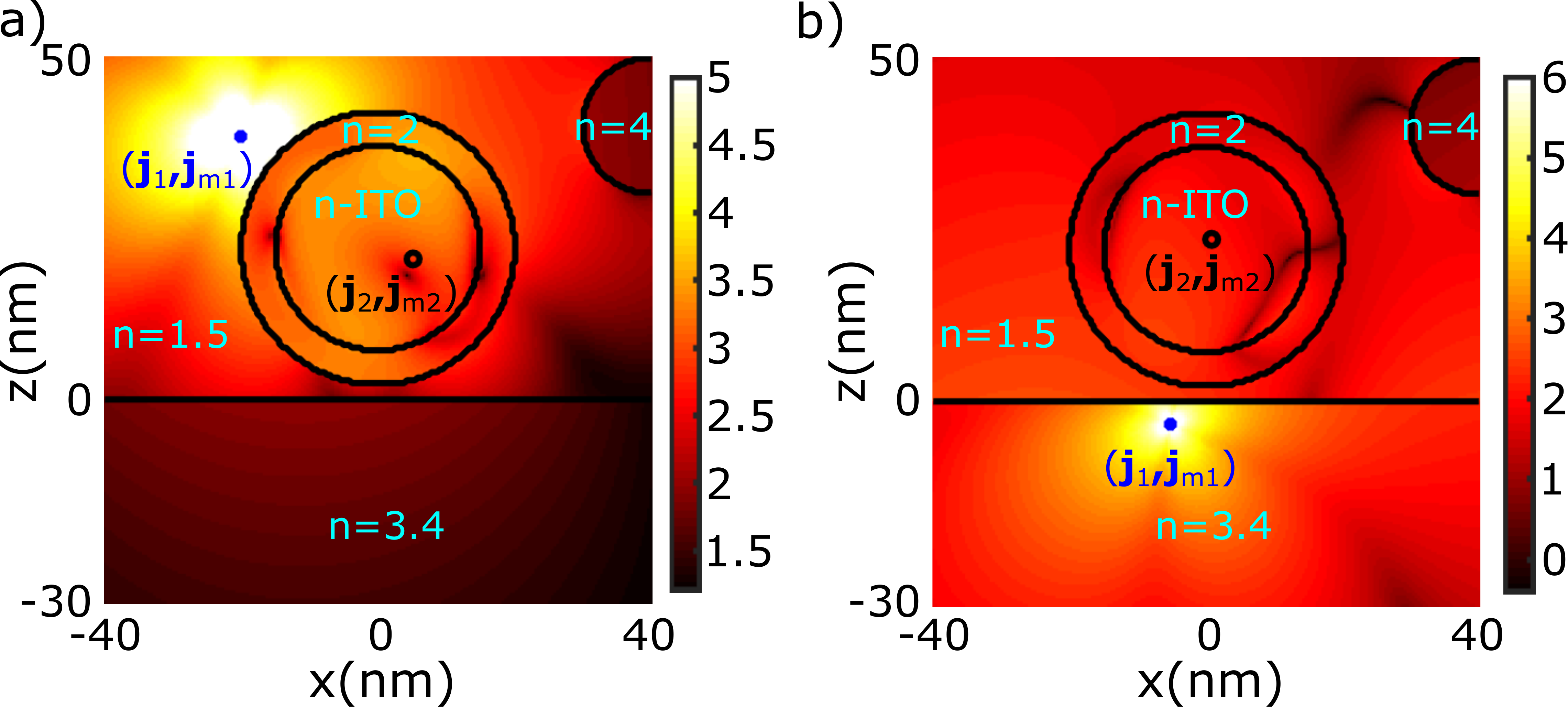}}
	\caption{Calculations performed with a multipole method \cite{Bertrand2020}. The electromagnetic field maps are plotted in log scale at $\lambda$=2.4$\mathrm{\mu}$m. Non-local reciprocity theorem verification with a complex electromagnetic environment composed of a 20-nm radius core-shell with a core of the same non-local medium (realistic permittivity of n-ITO) as in Fig. \ref{fig:champ} and a ligands shell of refractive index n=2 and 5-nm thickness. This core-shell is deposited on a silicon substrate (n=3.4) and is embedded in a background of n=1.5. In the vicinity of the core-shell is placed another nanosphere of refractive index n=4. In this complex environment, two point source currents ($\mathbf{j}_1,\mathbf{j}_{m1}$) and ($\mathbf{j}_2,\mathbf{j}_{m2}$) are inserted randomly. The reciprocity theorem of Eq.\ref{eq:rec_NL} is verified with 16 digits of precision.}
	\label{fig:champ_rec}
\end{figure}

This derivation has been done for the case where one source is outside and one source is inside the non-local medium but it is still valid if the two sources are in the same medium, whether it is local or not. We underline that placing a point source infinitely far is a straightforward way to generalize this non-local reciprocity theorem for a plane wave and a point source. Finally, it should be stressed that the reciprocity theorem we have derived gives the same expression as the classical theorem Eq. \ref{eq:rec} when the two points considered are located in the surrounding local medium, which is not necessarily obvious in presence of non-local media.

We have numerically checked that this non-local reciprocity theorem is satisfied for a complex setup including a non-local core-shell nanostructure (see Fig. \ref{fig:champ_rec}). We place the sources randomly, one inside the non-local medium, and the other one outside. The electromagnetic field maps created by these sources are then computed using a multipole method \cite{Bertrand2020}. The reciprocity theorem of Eq.\ref{eq:rec_NL} is verified with 16 digits of precision regardless of the position or polarization of the two sources.\\

\section{Conclusion}
We have derived here rigorous formulas which allow to generalize the classical expressions for the absorption volume density and the reciprocity theorem beyond the Drude model, to the case where non-local media described by the hydrodynamic model are present. Our numerical simulations show that the absorption volume density can significantly differ from classical predictions when nonlocality is taken into account. We underline also that our results can prove particularly useful to check that numerical methods based on the hydrodynamic model \cite{Toscano2012,schmitt2016dgtd} are indeed accurate, as done in the present work.

For a long time, nonlocality has been expected to play a role only in the tiniest metallic nanoparticles \cite{scholl2012quantum}. Recent results point towards a larger influence of nonlocality than previously expected, for much larger structures \cite{Pitelet2018,Pitelet2019}, for propagating surface plasmon \cite{Raza2013a}, and for semiconductors even more than for metals \cite{Maack2017,golestanizadeh2019hydrodynamic}. Moreover, the recent trend towards miniaturized plasmonic devices \cite{akselrod2014probing,haffner2015all} means that nonlocality will have, increasingly often, to be taken into account. Given the large number of situations in which local absorption plays a crucial role (photothermal therapy, local chemical reaction catalysis, HAMR,... etc.) we think our work could simply allow to better understand and finally to help design plasmonic nanostructures. Our work paves also the way to extend this work to more elaborated models based on the same HDM equations but going beyond the hard-wall boundary conditions \cite{Toscano2015,Ciraci2016}.

\begin{acknowledgments}
This work was supported by French state funds managed by the ANR within the Investissements d'Avenir programmes under reference ANR-20-CE24-0024-01 and 16-IDEX-0001 CAP 20-25. The authors thanks also Philippe Lalanne for the fruitful discussions at the beginning of this project. Antoine Moreau is an Academy CAP 20-25 chair holder.
\end{acknowledgments}
\appendix
\section{Absorption}
In this section, we show the derivation allowing to obtain Eq. \ref{eq:Formule2} of the main text. By using Eq. \ref{eq:new_eq5}, Eq. \ref{eq:Formule} of the main text can be written as a function of $\mathbf{P}_f$:
\begin{align*}
P_{\mathrm{abs}}(\omega) &
=\frac{\omega\varepsilon_0}{2}~\mathrm{Im}\left[\int_{V_e} \varepsilon|\textbf{E}|^2-\varepsilon\textbf{E}^*\textbf{e}\right] 
\\& = \frac{\omega\varepsilon_0}{2}~\mathrm{Im}\left[\int_{V_e} (\varepsilon-\chi_f)|\textbf{E}|^2+\textbf{E}^*\frac{\mathbf{P}_f}{\varepsilon_0}\right] 
\\& = 
\begin{multlined}[t]
\frac{\omega\varepsilon_0}{2}~\mathrm{Im}\left[\int_{V_e} (\varepsilon-\chi_f)|\textbf{E}|^2
+\frac{1}{\chi_f^*}\left|\frac{\mathbf{P}_f}{\varepsilon_0}\right|^2\right.\\
\left.+\frac{\varepsilon^*\textbf{e}^*}{\chi_f^*}\frac{\mathbf{P}_f}{\varepsilon_0}\right]
\end{multlined}
\end{align*}

The third term of this last equation can be expressed thanks to the Ostrogradski theorem and the hard-wall boundary conditions:
\begin{align}\label{eq:div_rhoPf}
\oiint_{S(V_e)} \rho^*(\frac{\textbf{P}_f}{\varepsilon_0}.\textbf{n})= & 0 =\int_{V_e}\nabla.(\rho^*\frac{\textbf{P}_f}{\varepsilon_0})
\\&0 =\int_{V_e}\textbf{e}^*\frac{\textbf{P}_f}{\varepsilon_0}+\rho^*\nabla.\frac{\textbf{P}_f}{\varepsilon_0}
\end{align}

Then by using Eq. \ref{eq:new_eq5} along with Eq. \ref{eq:disp_rho} and by expressing $\alpha$:
\begin{align*}
\nabla.\frac{\textbf{P}_f}{\varepsilon_0}&=\chi_f\nabla.\textbf{E}_0-(\varepsilon-\chi_f)\nabla.\textbf{e}
\\& =-(\varepsilon-\chi_f)\Delta.\rho
\\& =-(\varepsilon-\chi_f)\frac{\rho}{\alpha}
\\& =-\frac{\varepsilon}{\chi_f}\frac{\omega_p^2}{\beta^2}\rho
\end{align*}

which leads to the final expression of the absorption power inside $V_e$ corresponding to Eq. \ref{eq:Formule2} of the main text:
\begin{align*}\label{eq:Formule2_app}
P_{\mathrm{abs}}(\omega) &=
\begin{multlined}[t]
\frac{\omega\varepsilon_0}{2}~\mathrm{Im}\left[\int_{V_e} (\varepsilon-\chi_f)|\textbf{E}|^2+\frac{1}{\chi_f^*}\left|\frac{\mathbf{P}_f}{\varepsilon_0}\right|^2\right.\\
\left.+\left|\frac{\varepsilon}{\chi_f}\right|^2\frac{\omega_p^2}{\beta^2}|\rho|^2\right] 
\end{multlined}
\\& = \frac{\omega\varepsilon_0}{2}~\mathrm{Im}\left[\int_{V_e} (\varepsilon-\chi_f)|\textbf{E}|^2+\frac{1}{\chi_f^*}\left|\frac{\mathbf{P}_f}{\varepsilon_0}\right|^2\right] 
\\& =\frac{\omega\varepsilon_0}{2}~\int_{V_e}\left(\mathrm{Im}(\varepsilon-\chi_f)|\textbf{E}|^2+\mathrm{Im}(\chi_f)\left|\frac{\mathbf{P}_f}{\varepsilon_0\chi_f}\right|^2\right)
\end{align*}

\section{Longitudinal and transverse field intensity enhancement}

To help the reader interpreting the results of Fig. \ref{fig:champ}, we plot here the field intensity enhancement maps for the longitudinal and the transverse electromagnetic fields for the same parameters as the one used in Fig. \ref{fig:champ}.

\begin{figure}[h!]
    \centerline{\includegraphics[width=0.8\columnwidth]{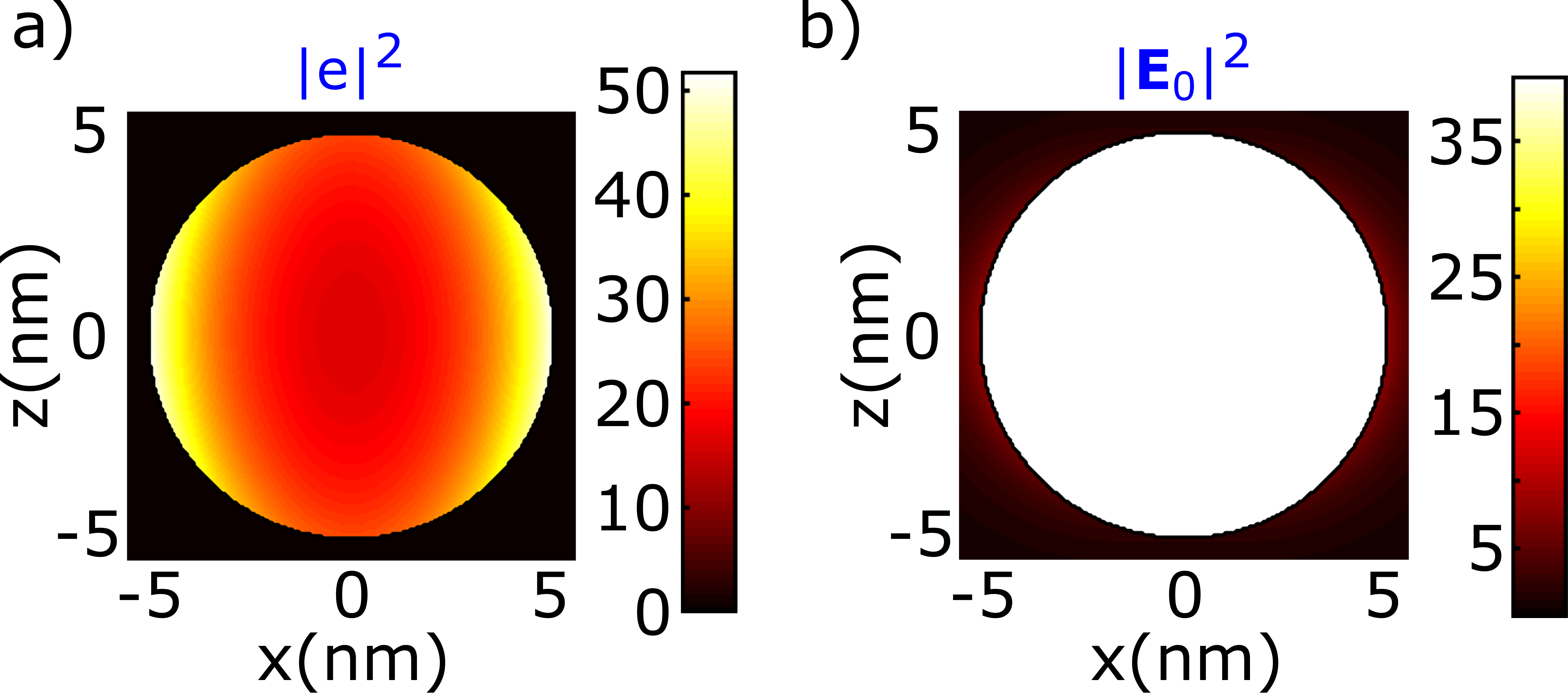}}
	\caption{Longitudinal and transverse field intensity enhancement for a sphere with the same parameters of materials and illumination conditions as Fig. \ref{fig:champ}.}
	\label{fig:Champ_app}
\end{figure}

\section{Absorption cross-section spectra of Fig. 4}
To complete the results presented in Fig. \ref{fig:champ_bis}, we plot here the absorption cross-section spectra of the two particles for the different cases treated in Fig. \ref{fig:champ_bis}. Let's precise here that the calculation of these spectra can be done equivalently with Eq. \ref{eq:Formule}, Eq. \ref{eq:Formule2} or the surface integral of the Poynting flux (we verified numerically that the spectra obtained with these three formula matches perfectly, similarly to what is shown on Fig. \ref{fig:spectre}). We retrieve once again the typical blue-shift of the non-local effects, that is obviously more important for the 5-nm-radius sphere than for the 25-nm-radius sphere.

\begin{figure}[h!]
    \centerline{\includegraphics[width=\columnwidth]{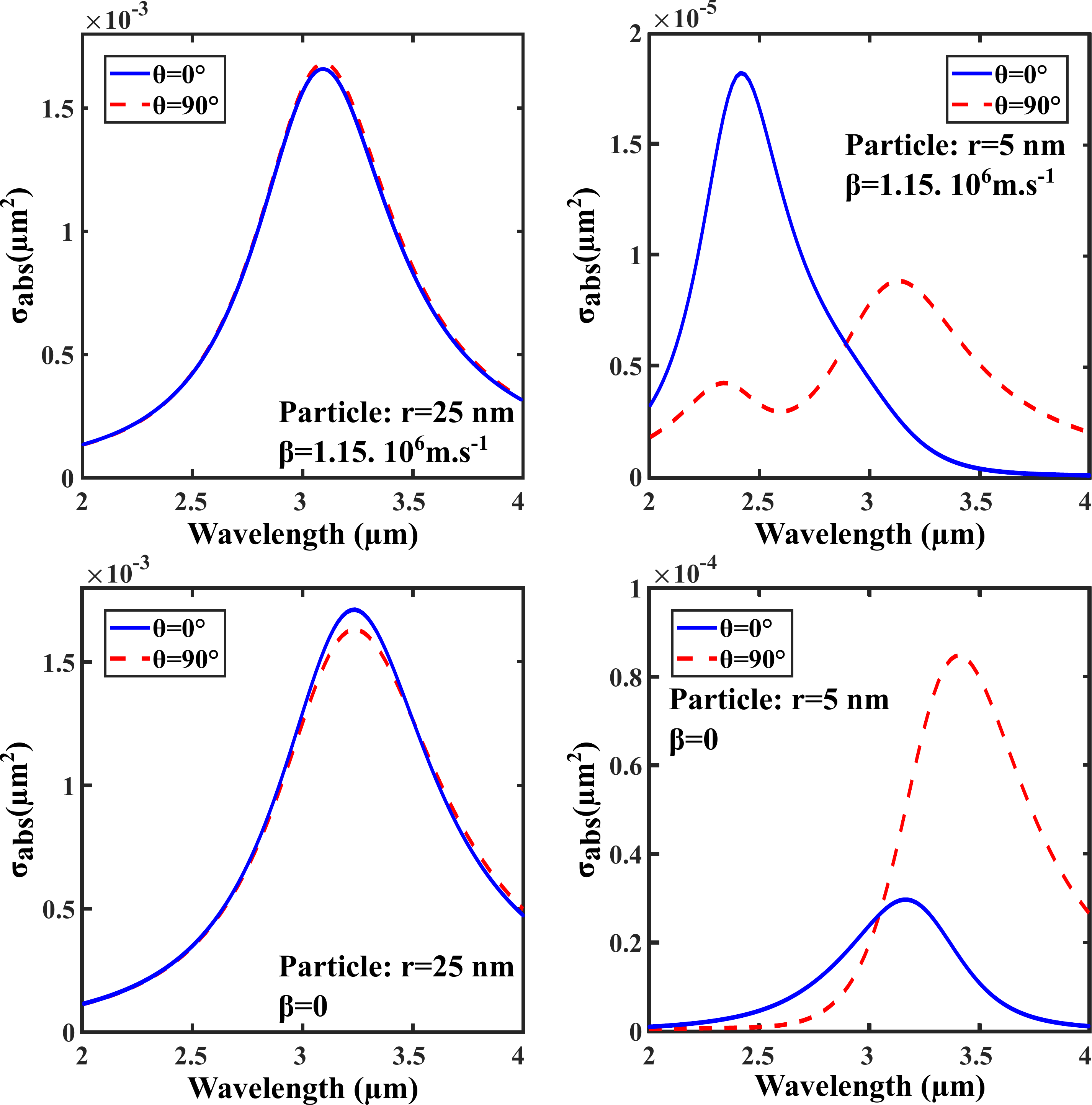}}
	\caption{Absorption cross-section spectra of the two particles for each cases treated in Fig. \ref{fig:champ_bis}. $\theta=0^{\circ}$  indicates the case where the nanoabsorber (5-nm-radius nanosphere) and the nanoantenna (25-nm-radius nanosphere) are aligned along the z-axis. $\theta=90^{\circ}$ indicates the case where they are aligned along the x-axis.}
	\label{fig:JP}
\end{figure}

\section{Reciprocity}
Eq. \eqref{eq:int_annex} relates two surface integrals, one outside and one inside the non-local medium. For the sake of completeness, we give here the derivation leading to this relation. In particular, thanks to the continuity of the tangential components of $\mathbf{E}$ and $\mathbf{H}$, it is possible to write the following expression:
\begin{equation}\label{eq:int_annex_tot}
\oiint_{S^+}(\mathbf{E}_1\times\mathbf{H}_2-\mathbf{E}_2\times\mathbf{H}_1).dS\mathbf{n} =
\oiint_{S^-}(\mathbf{E}_{1}\times\mathbf{H}_2-\mathbf{E}_{2}\times\mathbf{H}_1).dS\mathbf{n} 
\end{equation}
By definition of the field inside the non-local medium, the second term of this equation can also be written as:
\begin{equation}\label{eq:int_annex_2}
\oiint_{S^-}(\mathbf{E}_{01}\times\mathbf{H}_2-\mathbf{E}_{02}\times\mathbf{H}_1).dS\mathbf{n}+\oiint_{S^-}(\nabla\rho_{1}\times\mathbf{H}_2-\nabla\rho_{2}\times\mathbf{H}_1).dS\mathbf{n}
\end{equation}

To establish \eqref{eq:int_annex}, we demonstrate first that the second term of Eq. \eqref{eq:int_annex_2} is null, by noticing that

\begin{align*}
\nabla\rho_{1}\times\mathbf{H}_2&=\nabla\times(\rho_{1}\mathbf{H}_2)-\rho_{1}.\nabla\times\mathbf{H}_2\\
&=\nabla\times(\rho_{1}\mathbf{H}_2)-\rho_{1}.\left(-i\omega\varepsilon_0\varepsilon\mathbf{E}_{02}+\mathbf{j}_2\delta(\mathbf{r}-\mathbf{r}_2)\right).
\end{align*}

Symmetrically, we can write
\begin{align*}
\nabla\rho_{2}\times\mathbf{H}_1&=\nabla\times(\rho_{2}\mathbf{H}_1)-\rho_{2}.\left(-i\omega\varepsilon_0\varepsilon\mathbf{E}_{01}+\mathbf{j}_1\delta(\mathbf{r}-\mathbf{r}_1)\right).
\end{align*}

Using the Ostrogradski theorem then yields
\begin{equation*}
\oiint_{S^-}\nabla\times(\rho_{i}\mathbf{H}_j).dS\mathbf{n}=\int_{V_e}\nabla.\left(\nabla\times(\rho_{i}\mathbf{H}_j)\right)dV=0,
\end{equation*}
so that the second term in Eq. \eqref{eq:int_annex_2} can be rewritten as
\begin{equation}\label{eq:int_annex_0bis}
I=\oiint_{S^-}i\omega\varepsilon_0\varepsilon\left(\rho_{1}\mathbf{E}_{02}-\rho_{2}\mathbf{E}_{01}\right).dS\mathbf{n}.
\end{equation}

Then, using Eq. \eqref{eq:grad_rho} we get
\begin{equation}\label{eq:int_annex_rho}
I=\oiint_{S^-}i\omega\varepsilon_0\varepsilon\frac{(\varepsilon-\chi_f)}{\chi_f}\left(\rho_{1}\nabla\mathbf{\rho}_2-\rho_{2}\nabla\mathbf{\rho}_1\right).dS\mathbf{n},
\end{equation}
and by applying once again the Ostrogradski theorem, we obtain
\begin{align*}
I&=i\omega\varepsilon_0\varepsilon\frac{(\varepsilon-\chi_f)}{\chi_f}\int_{V_e}\nabla.\left(\rho_{1}\nabla\mathbf{\rho}_2-\rho_{2}\nabla\mathbf{\rho}_1\right).dV\\
&=i\omega\varepsilon_0\varepsilon\frac{(\varepsilon-\chi_f)}{\chi_f}\int_{V_e}\left(\rho_{1}\Delta\mathbf{\rho}_2-\rho_{2}\Delta\mathbf{\rho}_1\right).dV.
\end{align*}

Finally, using Eq. \eqref{eq:disp_rho} we obtain
\begin{equation}\label{eq:I_0}
I=i\omega\varepsilon_0\varepsilon\frac{(\varepsilon-\chi_f)}{\chi_f}\int_{V_e}\left(\rho_{1}\frac{\mathbf{\rho}_2}{\alpha}-\rho_{2}\frac{\mathbf{\rho}_1}{\alpha}\right).dV=0,
\end{equation}
which establishes the continuity of the surface integral described by Eq. \eqref{eq:int_annex}.

\end{document}